\documentclass[aps,prd,secnumarabic,amssymb,amsmath,nobibnotes,nofootinbib,notitlepage]{revtex4-1}

\usepackage{amsfonts,amsmath,hyperref,url, color}
\usepackage{eurosym}
\usepackage{graphicx}

\usepackage{bbm}
\usepackage{slashed}
\usepackage{braket}

\usepackage{ulem}

\newcommand{\id}{\mathbbm{1}}
\newcommand{\p}{\slashed{p}}
\newcommand{\so}{\mathfrak{so}}
\newcommand{\su}{\mathfrak{su}}
\newcommand{\g}{\mathfrak{g}}

\renewcommand{\sl}{\mathfrak{sl}}

\newcommand{\C}{\mathbb{C}}

\begin{document}

\title{Path integral quantization of a spinning particle }

\author{Jerzy Kowalski-Glikman}
\email{jerzy.kowalski-glikman@uwr.edu.pl}
\affiliation{Institute for Theoretical Physics, University of Wroc\l{}aw, pl.\ M.\ Borna 9, 50-204 Wroc\l{}aw, Poland}
\affiliation{National Centre for Nuclear Research, Pasteura 7, 02-093 Warsaw, Poland}

\author{Giacomo Rosati}
\email{giacomo.rosati@uwr.edu.pl}
\affiliation{Institute for Theoretical Physics, University of Wroc\l{}aw, pl.\ M.\ Borna 9, 50-204 Wroc\l{}aw, Poland}

\begin{abstract}
Following the idea of Alekseev and Shatashvili we derive the path integral quantization of a modified relativistic particle action that results in the Feynman propagator of a free field with arbitrary spin. This propagator can be associated with the Duffin, Kemmer, and Petiau (DKP) form of a free field theory. We show explicitly that the obtained DKP propagator is equivalent to the standard one, for spins 0 and 1. We argue that this equivalence holds also for higher spins.
\end{abstract}

\maketitle

\section{Introduction}

Spin is a subtle and elusive concept. On the one hand one would think that it should be very easy to describe. Indeed, like the momentum carried by a particle is an eigenvalue of the translation operator, spin can be described as an eigenvalue of the operators associated with rotations (usually taken to be the $z$-component of angular momentum and the angular momentum squared). One would think that it is pretty easy then to find a description of a classical relativistic particle with spin and then to quantize it, so as to obtain the free theory of a field with spin. But this is not the case. There are many attempts to formulate a classical spin theory, both in terms of commuting  and anticommuting classical variables, that after quantization produce the expected quantum fields with spin, reviewed, for example in~\cite{Frydryszak:1996mu}. Among the commutative models a particularly interesting one was formulated by Balachandran et. al.~\cite{Balachandran:1979ha},~\cite{Balachandran:2017jha}, who assumed that the configuration space of a spinning particle should be identified with the Poincar\'e group. This construction was later found very fruitful, for example in the case of particles coupled to gravity in three~\cite{deSousaGerbert:1990yp} and four~\cite{Freidel:2006hv} spacetime dimensions. Some other approaches are reviewed in the recent paper~\cite{Rempel:2016jbn}.

A particularly convenient approach to quantization is path integral. It is well known that when one uses the path integral to describe the quantum transition amplitude of a relativistic particle one gets as a result a scalar (Feynman) propagator of the form $(p^2-m^2 +i\epsilon)^{-1}$. This is perfectly consistent with the result of canonical quantization, in which case the classical first-class constraint $p^2 - m^2 =0$ becomes, after quantization, according to Dirac procedure, the Klein-Gordon equation $(\square+m^2)\phi =0$.

The question arises if one can find the `spinning' relativistic particle action such that, after plugging it into the path integral, one gets as a result the Dirac propagator $(\p-m+i\epsilon)^{-1}$ in the case of spin $1/2$, and appropriate expressions for higher spins. Clearly, this requires two kinds of modifications of the standard relativistic particle action. First, contrary to the scalar propagator, Dirac propagator has a term linear in momentum. Second, the path integral should produce the right representation of Dirac $\gamma$-matrices from some classical data. It is the aim of the present paper to describe such construction in details. In our approach we follow the scheme proposed some time ago by Alekseev and Shatashvili \cite{Alekseev:1988tj}, which, in turn, was motivated by the construction proposed by Polyakov in \cite{Polyakov:1988md}.

To this aim in Section II, after presenting the standard scalar path integral, we observe that the Dirac form of the propagator can be obtained if we start with a relativistic particle action in which the first-class constraint becomes linear in momenta. Parallelly to that, we notice that an analogous approach can be phrased in terms of the Duffin, Kemmer, and Petiau (DKP) formalism for a field theory with spin 0 and 1. We first show how , for a scalar (spin 0) field, the second order formalism can be naturally associated with the first order DKP formulation. In Section III we construct the spinning particle path integral, obtaining the general expression for the propagator. This section relies on some more technical results, which are described in details in  Appendices. In the next section IV we complete the explicit construction of the propagator for spin 0, 1/2, and 1, and we prove that the so obtained DKP propagators are equivalent to the standard ones. We do not attempt to extend our construction to spins higher than 1 explicitly, although there are little doubts that such a generalization can be done. Unfortunately, the technical difficulty of the formalism grows rapidly with growing spin, as the equations defining the higher spin analogues of Dirac $\gamma$-matrices are getting more and more complicated. The final Section V is devoted to conclusions and discussion of open problems.

\section{The scalar path integral}

Consider the standard, free, scalar relativistic particle moving in
four-dimensional spacetime, between the spacetime point with coordinates
$x_{1}$ and the one with coordinates $x_{2}$. The  transition amplitude from the initial state $\left<\mbox{in}\right|= \left<x_1\right|$ to the final one $\left|\mbox{out}\right>= \left|x_2\right>$ is given by the path integral\footnote{In this manuscript we denote the four dimensional indices by $\mu=0,1,2,3,$, raised and lowered
by the 4D  Minkowski metric $\eta_{\mu\nu}=\text{diag}\left(1,-1,-1,-1\right)$;
 $u\cdot v$
and $v^{2}$ are shorthand for $u^{\mu}v_{\mu}$ and $v_{\mu}v^{\mu}$ respectively,
while $\boldsymbol{u}\cdot\boldsymbol{v}=\sum_{i=1}^{3}u_{i}v_{i}$.}
for the trajectories beginning at $x_{1}$ and ending in $x_{2}$:
\begin{equation}
G(x_{2},x_{1})=\int_{x(0)=x_{1}}^{x(1)=x_{2}}{\cal D}(x(\tau)){\cal D}(p(\tau)){\cal D}({\cal N}(\tau)/{\rm Diff})\,e^{iS},
\label{Green}
\end{equation}
where the action $S$ is
\begin{equation}
S=\int_{0}^{1}d\tau\left(p\dot{x}-{\cal N}(p^{2}-m^{2})\right),
\label{ActionClassic}
\end{equation}
and ${\cal D}({\cal N}(\tau)/{\rm Diff})$ denotes the measure on
the Lagrange multiplier ${\cal N}$ up to worldline reparametrization,
under which it transforms as a one-dimensional metric determinant.

Now we integrate over $x(\tau)$. In order to do that we must first rewrite
it in the form that conveniently takes into account the boundary conditions:
\begin{equation}
x(\tau)=x_{1}+(x_{2}-x_{1})\tau+y(\tau)\,,\quad y(0)=y(1)=0.\label{BoundaryCondition}
\end{equation}
Clearly ${\cal D}(x(\tau))={\cal D}(y(\tau))$. Now we can integrate
by parts the action (\ref{ActionClassic}), and then integrate over
$y(\tau)$ obtaining
\begin{equation}
G(x_{2},x_{1})=\int{\cal D}(p(\tau)){\cal D}({\cal N}(\tau)/{\rm Diff})\,\delta(\dot{p})\,e^{ip(x_{2}-x_{1})}\,\exp\left(-i(p^{2}-m^{2})\,\int_{0}^{1}d\tau{\cal N}\right)\label{Green2}
\end{equation}
Noticing now that $\delta(\dot{p})$ enforces the momenta to be $\tau$-independent,
so that ${\cal D}(p(\tau))\delta(\dot{p})=d^{4}p$, and that
\[
L\equiv\int_{0}^{1}d\tau{\cal N}>0
\]
is the gauge invariant information carried by ${\cal N}$\footnote{It is essential at this point that ${\cal N}$, being the one dimensional
Euclidean metric is positive. See \cite{Freidel:2018apz} for the recent detailed discussion of this issue.}, we can express the path integral (\ref{Green2}) as
\begin{equation}
G(x_{2},x_{1})=\int d^{4}p\,\int_{0}^{\infty}dL\,e^{ip(x_{2}-x_{1})}\,e^{-iL(p^{2}-m^{2}-i\epsilon)}=\int d^{4}p\,\frac{i}{p^{2}-m^{2}-i\epsilon}\,e^{ip(x_{2}-x_{1})}\label{propagatorScalar}
\end{equation}
where we added the $-i\epsilon$ term to regularize the integral,
as usual. The Fourier transform of the transition amplitude (\ref{propagatorScalar}) is the (Feynman) propagator of the quantum scalar field.
\newline

The approach outlined above cannot be directly applied to the case of fields with higher spin. For example, in the case of spin 1/2 field the propagator is the inverse of an expression linear in momentum, $(\slashed{p}-m)^{-1}$, instead of the inverse of quadratic expression, $(p^2-m^2)^{-1}$, as in (\ref{propagatorScalar}). It was not long after Dirac's formulation of a theory of spin $1/2$-fields, when a similar (unified) formulation for fields of spin-0 and spin-1 has been put forward by Duffin, Kemmer, and Petiau (DKP) \cite{Duffin:1938zz,kemmerDKP,Petiau}. While the details of the DKP theory needed for our analysis will be discussed in Sec.~\ref{sec:DKPpropagator}, let us present here a brief introduction to this approach for the scalar (spin-0) fields.

The very reason behind the $p^2-m^2$ term in the scalar field propagator is the form of scalar field equations that follow from the lagrangian
\begin{equation}\label{sflag}
  L = \partial_\mu\phi\partial^\mu\phi - m^2\phi^2
\end{equation}
In order to get the DKP propagator, inverse proportional to momentum (instead of its square), we must rewrite the lagrangian (\ref{sflag}) in the form linear in space-time derivatives, similar to the form of Dirac lagrangian. This can be achieved by turning from the second order formulation (with second order derivatives) to the first order one, in which the field $\phi$ and its derivatives $\partial \phi$ are treated as independent field components of a multi-component field
\begin{equation}\label{psi}
   \psi =(\pi_0, \pi_1,\pi_2,\pi_3,\varphi)^T .
\end{equation}

For a real scalar field $\varphi^\dag=\varphi$ the DKP lagrangian takes the form~\footnote{ Here we adopt a specific representation for the $\beta$ matrices presented in the Appendix~\ref{sec:DKP}, and we refer to Sec.~\ref{sec:DKPpropagator} for details.}
\begin{equation}\label{DKPlag}
\begin{split}
L_{DKP} & = \frac{i}{2} \bar\psi\beta^\mu\partial_\mu\psi - \frac{i}{2} (\partial_\mu\bar\psi)\beta^\mu\psi - m\bar\psi\psi\,\\ &
= \frac{i}{2}\sum_{\mu}\left(\left(\pi_{\mu}^{\dagger}-\pi_{\mu}\right)\partial_{\mu}\varphi-
\varphi\partial_{\mu}\left(\pi_{\mu}^{\dagger}-\pi_{\mu}\right)\right)-m\pi_{\mu}^{\dagger}\pi^{\mu}-m\varphi^{2} ,
\end{split}
\end{equation}
where $\beta^\mu$ are the so-called DKP $\beta$-matrices, playing, for the spin 0 and spin 1 theories, a role analogous to that of Dirac $\gamma$ matrices for spin 1/2, and the adjoint field $\bar\psi$ are defined as
\begin{equation}\label{psiBar}
   \bar\psi =\psi^\dag\, \eta_0=(\pi^\dag_0, -\pi^\dag_1,-\pi^\dag_2,-\pi^\dag_3,\varphi^\dag)
\end{equation}
with $\eta_0\equiv 2\beta_0^2-\eta_{00}\mathbbm{1}$.

Varying the DKP Lagrangian with respect to $\pi_\mu$ and $\pi_\mu^\dagger$, we obtain the expression for the conjugate momenta
\begin{equation}
\pi_\mu = \frac{i}{m}\eta^{\mu\nu}\partial_{\nu}\varphi,\qquad \pi_{\mu}^{\dagger}=-\frac{i}{m}\eta^{\mu\nu}\partial_{\nu}\varphi,
\end{equation}
which, substituted into the DKP lagrangian, gives back, after the identification $\varphi=\sqrt m \phi$, the quadratic lagrangian~(\ref{sflag}).
This shows that the two lagrangians are equivalent (both classically and quantum mechanically), and one concludes that, for free fields,
the DKP formalism is nothing but using the  first order lagrangian.

It follows  that the DKP Lagrangian leads to the quantum propagator of the form
\begin{equation}\label{dkfpropagator}
 G(p) = \frac1{\slashed{p} - m\mathbbm{1}-i\epsilon}\,, \quad \slashed{p} =\beta_\mu p^\mu
\end{equation}
It is expected that an analogous construction can be made for higher spins. In that case the propagators for higher spins will have the same form, but with appropriately chosen matrices replacing the $\beta$ matrices of spin-0/spin-1 theory.

A natural question arises as if it is possible to obtain this propagator from the path integral with some form of the particle action, as it was in the case of the scalar field above, (\ref{propagatorScalar}). The answer is positive, and in the next section we present the explicit construction.

\section{The spinning particle path integral}
\label{sec:spinning}

In this section we will discuss how the path integral for spinning
particle can be written in the form proposed by Alekseev and Shatashvili~\cite{Alekseev:1988tj}, whose construction is, in turn, a generalization of the one of Polyakov~\cite{Polyakov:1988md}.
We will omit the more technical aspects of the argument, presenting
them in details in the Appendix, stressing here the motivations and
the meaning of the final result.

Our starting point is the path integral (\ref{Green2}) in
momentum representation
\begin{equation}
G(p)=\int\,{\cal D}({\cal N}(\tau)/{\rm Diff})\,\exp\left(-i\int_{0}^{1}d\tau{\cal N}(p^{2}-m^{2})\right),\label{Gp}
\end{equation}
Our goal is to generalize the form of (\ref{Gp}) so as to make it describe  a particle of an arbitrary spin.

We start with the observation that the action in (\ref{Gp}) can be
rewritten as
\begin{equation}
{\cal N}(p^{2}-m^{2})=m{\cal N}(p\cdot p/m-m)={\cal N}'(p\upsilon-m)\label{actionalpha}
\end{equation}
where we introduced a new variable $\upsilon$ that replaces $p/m$. The variable $\upsilon$ is, of course, nothing but the four velocity, satisfying $\upsilon^2=1$ and therefore belonging to the 3-dimensional pseudo-sphere $PS^3$.

Let us stop for a moment to contemplate on the meaning of \eqref{actionalpha}. We replaced the second order constraint of the particle action $p^{2}-m^{2}=0$ with the first order one $p\upsilon-m=0$. The former leads to the standard scalar propagator \eqref{propagatorScalar}, and it is natural to expect that the latter will lead to the DKP one \eqref{dkfpropagator} if we force the path integral to replace $\upsilon$ with the DKP matrices $\beta$. Now, the $\beta$ matrices, similarly to the Dirac matrices, are defined to satisfy (among others) the requirement that their commutator has the form
\begin{equation}\label{betacommutator}
  [\beta_\mu, \beta_\nu]= S_{\mu\nu},
\end{equation}
where $S_{\mu\nu}$ generate Lorentz transformations~\cite{LunardiGauge} $U \simeq 1 + \tfrac{1}{2}\omega^{\mu\nu}S_{\mu\nu}$.

Since the commutator \eqref{betacommutator} must come as a result of quantization of a classical theory, the kinetic term (symplectic form) of the latter should be such that the associated Poisson bracket has the form
\begin{equation}\label{upsilonpoisson}
  \{\upsilon_\mu, \upsilon_\nu\}=  j_{\mu\nu},
\end{equation}
where, again, $j$ are Lorentz generators, satisfying $\so(3,1)$ algebra.
After quantization (as we will show below), the associated quantum operators satisfy the commutators
\begin{equation}\label{upsilonComm}
  [\hat{\upsilon}_\mu, \hat{\upsilon}_\nu ]=  i \hat{j}_{\mu\nu},
\end{equation}
and one gets (\ref{betacommutator}) after identifying
\begin{equation}
\beta_\mu \equiv \hat{\upsilon}_\mu, \qquad S_{\mu\nu} \equiv \hat{S}_{\mu\nu} = i \hat{j}_{\mu\nu}.
\label{mapSigmaS}
\end{equation}
It turns out that in order to get the correct properties for the $\beta$ (as well as for the Dirac matrices $\gamma$, see for instance~\cite{WeinbergBook} Ch. 5.4), the operators $\hat{v}_\mu$ must be generators of Clifford algebra $Cl_{3,1}$. Their operators $\hat{j}_{\mu\nu}$ generate the $\so(3,1)$ Lorentz algebra, and one can show that together, $\hat{v}_\mu$ and $\hat{j}_{\mu\nu}$ generate the $\so(3,2)$ Lie algebra, the anti de Sitter algebra.
In turn, the matrices $\beta_\mu$ (or $\gamma_\mu$) and $S_{\mu\nu}$, obtained by the substitution (\ref{mapSigmaS}) generate the $\so(4,1)$ de Sitter algebra.

Now, since in the Poisson-Lie theory there is a one-to-one correspondence between commutators of the algebra and the Poisson structure on the dual algebra, it is natural to to identify $\upsilon$ with elements of the Lie algebra $\so(3,2)^*$ ``dual'' to the one spanned by the generators of $\so(3,2)$. Let us discuss in details how this comes about.

We start from the $\so(3,2)$ Lie algebra, generated by the antisymmetric matrices ($A,B=0,1,2,3,4$)
\begin{equation}
(M_{AB})^C_{\ D}=-(M_{BA})^C_{\ D} = \delta^C_A \eta_{BE}-\delta^C_B \eta_{AE},\qquad \eta_{AB} = \text{diag}(1,-1,-1,-1,1).
\end{equation}
and defined by the Lie brackets
\begin{equation}
\left[M_{AB},M_{CD}\right]=\eta_{AD}M_{BC}+\eta_{BC}M_{AD}-\eta_{AC}M_{BD}-\eta_{BD}M_{AC}.\label{so(3,2)}
\end{equation}
With the redefinition $\Upsilon_\mu = M_{\mu4}$, $J_{\mu\nu} = M_{\mu\nu}$ ($\mu,\nu = 0,1,2,3$), the $\so(3,2)$ algebra takes the form
\begin{equation}
\begin{gathered}\left[ \Upsilon_{\mu},\Upsilon_{\nu}\right] =J_{\mu\nu},\qquad
\left[ J_{\mu\nu},\Upsilon_{\rho} \right] =\Upsilon_{\nu}\eta_{\mu\rho}-\Upsilon_{\mu}\eta_{\nu\rho},\\
\left[ J_{\mu\nu},J_{\rho\sigma}\right] =\eta_{\mu\rho}J_{\nu\sigma}+\eta_{\nu\sigma}J_{\mu\rho}-\eta_{\mu\sigma}J_{\nu\rho}-\eta_{\nu\rho}J_{\mu\sigma}.
\end{gathered}
\label{cl(3,1)algebra}
\end{equation}

An arbitrary element $X$ of the dual algebra $\so(3,2)^*$ is
spanned by the generators $\{\tilde{\Upsilon}^{\mu},\tilde{J}^{\mu\nu}\}$ ($\tilde{J}^{\nu\mu}=-\tilde{j}^{\mu\nu}$), dual to $\{\Upsilon_{\mu},J_{\mu\nu}\}$ in the sense that (see Appendix
\ref{sec:ActionOrbit}) $\langle \tilde{\Upsilon}^{\mu} , \Upsilon_\nu \rangle = \delta^\mu_\nu$ , $\langle \tilde{J}^{\mu\nu} , J_{\rho\sigma} \rangle = \delta^\mu_\rho\delta^\nu_\sigma - \delta^\nu_\rho\delta^\mu_\sigma$ , $\langle \tilde{\Upsilon}^{\mu} , J_{\rho\sigma} \rangle =  \langle \tilde{J}^{\mu\nu} , \Upsilon_\nu \rangle = 0$ , and it
has the form ($j_{\nu\mu}=-j_{\mu\nu}$)
\begin{equation}
X= \upsilon_{\mu}\tilde{\Upsilon}^{\mu}+ \tfrac{1}{2}j_{\mu\nu}\tilde{J}^{\mu\nu}.
\end{equation}
Using the definitions presented in Appendix \ref{sec:ActionOrbit}
one can check that the coadjoint orbit of $\upsilon\equiv\upsilon_\mu\tilde{\Upsilon}^{\mu}$
under the action of the Lorentz subgroup is exactly the pseudosphere
$PS^{3}$.
The orbits are characterized by the values $\left(c_{1},c_{2}\right)$
of the two polynomials of $v_{\mu}$, $j_{\mu\nu}$ invariant under the coadjoint action
of SO(3,2), corresponding to the two Casimirs of $\so(3,2)$:
\begin{equation}
\begin{gathered}{\cal C}_{1}=\upsilon^{\mu}\upsilon_{\mu}+\frac{1}{2}j_{\mu\nu}j^{\mu\nu},\\
{\cal C}_{2}=W_{0}^{2}-W\cdot W+\frac{1}{4}\left(\epsilon_{ijk}j_{jk}j_{i0}\right)^{2},
\end{gathered}
\label{casimir}
\end{equation}
with
\begin{equation}
W_{0}=\frac{1}{2}\epsilon_{ijk}j_{jk}\upsilon_{i},\qquad W=-\frac{1}{2}\epsilon_{ijk}j_{jk}\upsilon_{0}+\epsilon_{ijk}j_{j0}\upsilon_{k}.
\end{equation}

The action in the path integral should therefore consist of two pieces. The first is given by \eqref{actionalpha}, and the second is an action $S$ to be defined so as to impose the condition \eqref{upsilonpoisson}, and which leads to its quantization. It is given by \cite{Alekseev:1988vx}
\begin{equation}
S=\int\omega\label{actionS}
\end{equation}
where $\omega=\langle X(g), dg g^{-1}\rangle$, $g$ being an element of SO(3,2),  is the Liouville form associated with  Kirillov
symplectic two-form, discussed in details in Appendix \ref{sec:ActionOrbit}.
This action leads to the following expressions for the Poisson brackets
of the dynamical variables
\begin{equation}
\begin{gathered}\left\{ \upsilon_{\mu},\upsilon_{\nu}\right\} =j_{\mu\nu},\qquad
\left\{ j_{\mu\nu},\upsilon_{\rho}\right\} =\upsilon_{\nu}\eta_{\mu\rho}-\upsilon_{\mu}\eta_{\nu\rho},\\
\left\{ j_{\mu\nu},j_{\rho\sigma}\right\} =\eta_{\mu\rho}j_{\nu\sigma}+\eta_{\nu\sigma}j_{\mu\rho}-\eta_{\mu\sigma}j_{\nu\rho}-\eta_{\nu\rho}j_{\mu\sigma}.
\end{gathered}
\label{PoissonBracketsso(4,1)}
\end{equation}
This is exactly what we want, because after quantization the first
equation above will become the defining equation for the $\beta$
matrices of the DKP formalism.

The final form of the momentum space propagator is therefore
\begin{equation}
G(p)=\int_{0}^{\infty}dL\,\int{\cal D}\left(\Upsilon(t),J(t)\right)\,\exp\left(imL-i\int_{0}^{L}dt\,p\cdot\upsilon\left(t\right)\right)\exp\left(iS\left(v,j\right)\right),\label{GreenSpin}
\end{equation}
\begin{equation}
S\left(v,j\right)=\int_{0}^{L}dt\ \omega\left(v(t),j(t)\right).\label{ActionOrbitSO(4,1)}
\end{equation}
It is shown in the Appendix (\ref{sec:OrbitsIrreps}) that
the term $\exp\left(iS\left(v,j\right)\right)$ `quantizes' the
values of the invariant polynomials in $\{v_{\mu},j_{\mu\nu}\}$ defining the orbits,
so that the corresponding operators $\{\hat{v}_{\mu},\hat{j}_{\mu\nu}\}$ belong to an irreducible representation
$\{\pi(\hat{\upsilon}_{\mu}),\pi(\hat{j}_{\mu\nu})\}$ of $\so(3,2)$ (or, through the substitution $\hat{S}_{\mu\nu} = i \hat{j}_{\mu\nu}$, to an irreducible representation $\{\pi(\hat{v}_{\mu}),\pi(\hat{S}_{\mu\nu})\}$ of $\so(4,1)$). In other words the path integral in \eqref{GreenSpin} computes, for given boundary conditions, the correlation
function
\begin{equation}
\int{\cal D}\left(v(t),j(t)\right)\,\exp\left(imL-i\int_{0}^{L}dt\,p\cdot\upsilon\left(t\right)\right)\exp\left(iS\left(v,j\right)\right)=\left<i\left|\exp\left(i\int_{0}^{L}\,p\cdot \hat{v} \ dt\right)\right|j\right>\label{correlation}
\end{equation}
between states $\left|i\right>$ belonging to a particular representation of the
$\so(4,1)$ algebra, corresponding to the particular choice of integral
orbit. In the formula \eqref{correlation} $\hat{v}_\mu$ is the quantum operator corresponding to $v_\mu$, and, depending on the spin representation $\left|i\right>$, it is given by a Dirac $\gamma$ matrix (for spin 1/2) or a DKP $\beta$ matrix for spin 0 or 1, and, presumably, to matrix representations for higher spins.  In particular, these matrices must satisfy \eqref{betacommutator}, and we will show now that this is indeed the case.

Let us denote by $l_a \equiv \{v_\mu,j_{\mu\nu}\}$ the coordinates on the dual Lie algebra $\so(3,2)^*$, so that the Poisson brackets (\ref{PoissonBracketsso(4,1)}) can be written concisely in terms of the $\so(3,2)$ structure constant $f_{ab}^{\ \ c}$ defined by (\ref{cl(3,1)algebra}) as $\{l_a,l_b\}=f_{ab}^{\ \ c}$.
The transformation
\begin{equation}
l_{a}\rightarrow l_{a}+f_{ab}^{\ \ c}\xi^{b}l_{c}\label{variationl}
\end{equation}
is a symmetry of the classical action, and the resulting Ward identity
reads (see Appendix \ref{sec:correlation} for details)
\begin{equation}
\partial_{t}\braket{i|\hat{l}_{a}\left(t\right)\hat{l}_{a_{1}}\left(t_{1}\right)\cdots \hat{l}_{a_{n}}\left(t_{n}\right)|j}=i\sum_{k}f_{aa_{k}}^{\ \ b}\braket{i|\hat{l}_{b}\left(t\right)\hat{l}_{a_{1}}\left(t_{1}\right)\cdots \hbox{\sout{$\hat{l}_{a_{k}}\left(t_{k}\right)$}}\cdots \hat{l}_{a_{n}}\left(t_{n}\right)|j}\delta\left(t-t_{k}\right),\label{ward}
\end{equation}
where \sout{$\hat{l}_{a_{k}}\left(t_{k}\right)$} indicates that the particular term is missing.

In order to derive the equal time commutators (ETC) for the corresponding
field operators we can apply the BJL (Bjorken-Johnson-Low) procedure
to the correlation function, stating that the $1/p_{0}$ term in the
matrix element of the two point function, at large $p_{0}$, determines
the commutator:
\begin{equation}
\lim_{p_{0}\rightarrow\infty}p_{0}\int dte^{ip_{0}\left(t-t_{1}\right)}\braket{i|\hat{l}_{a}\left(t\right)\hat{l}_{a_{1}}(t_{1})|j}=i\braket{i|\left[\hat{l}_{a},\hat{l}_{a_{1}}\right]\left(t_{1}\right)|j},\label{BJL}
\end{equation}
where $[\hat{l}_{a},\hat{l}_{a_{1}}]$ is the ETC between field operators
corresponding to $l_{a},l_{a_1}$.
Integrating the left hand side in the last
expression by parts we rewrite it as
\begin{equation}
i\lim_{p_{0}\rightarrow\infty}\int dte^{ip_{0}\left(t-t_{1}\right)}\frac{\partial}{\partial t}\braket{i|\hat{l}_{a}\left(t\right)\hat{l}_{a_{1}}\left(t_{1}\right)|j},
\end{equation}
where we neglected boundary terms. From (\ref{ward}) the last expression
is equal to
\begin{equation}
-\lim_{p_{0}\rightarrow\infty}\int dte^{ip_{0}\left(t-t_{1}\right)}f_{aa_{1}}^{\ \ b}\braket{i|\hat{l}_{b}\left(t\right)|j}\delta\left(t-t_{1}\right)=-\lim_{p_{0}\rightarrow\infty}f_{aa_{1}}^{\ \ b}\braket{i|\hat{l}_{b}\left(t_{1}\right)|j},
\end{equation}
so that from (\ref{BJL}) one gets
\begin{equation}
\left[\hat{l}_{a},\hat{l}_{b}\right]=if_{ab}^{\ \ c}\hat{l}_{c},\label{EqualTimeCommutator},
\end{equation}
or, expanding in terms of the operators $\hat{\upsilon},\hat{j}$,
\begin{equation}
\begin{gathered}\left[ \hat{\upsilon}_{\mu},\hat{\upsilon}_{\nu}\right] =i \hat{j}_{\mu\nu},\qquad
\left[ \hat{j}_{\mu\nu},\hat{\upsilon}_{\rho}\right] =i\left(\hat{\upsilon}_{\nu}\eta_{\mu\rho}-\hat{\upsilon}_{\mu}\eta_{\nu\rho}\right),\\
\left[ \hat{j}_{\mu\nu},\hat{j}_{\rho\sigma}\right] =i \left(\eta_{\mu\rho}\hat{j}_{\nu\sigma}+\eta_{\nu\sigma}\hat{j}_{\mu\rho}-\eta_{\mu\sigma}\hat{j}_{\nu\rho}-\eta_{\nu\rho}\hat{j}_{\mu\sigma}\right).
\end{gathered}
\label{so32commutators}
\end{equation}

Finally, using the substitution (\ref{mapSigmaS}), we can rewrite the commutators as
\begin{equation}
\begin{gathered}\left[ \hat{\upsilon}_{\mu},\hat{\upsilon}_{\nu}\right] = \hat{S}_{\mu\nu},\qquad
\left[ \hat{S}_{\mu\nu},\hat{\upsilon}_{\rho}\right] =\hat{\upsilon}_{\mu}\eta_{\nu\rho}-\hat{\upsilon}_{\nu}\eta_{\mu\rho},\\
\left[ \hat{S}_{\mu\nu},\hat{S}_{\rho\sigma}\right] =\eta_{\mu\sigma}\hat{S}_{\nu\rho}+\eta_{\nu\rho}\hat{S}_{\mu\sigma}-\eta_{\mu\rho}\hat{S}_{\nu\sigma}-\eta_{\nu\sigma}\hat{S}_{\mu\rho}.
\end{gathered}
\label{so41commutators}
\end{equation}
This is nothing but the Lie algebra (with real structure constants) $\so(4,1)$ of $SO(4,1)$, which proves that, after computing the path integral in the formula \eqref{correlation}, the operators $\hat{\upsilon}_\mu$ can be taken to be a matrix of a particular representation of the $\mathfrak{so}(4,1)$
algebra.
We will show in the next session how, depending on the specific $\mathfrak{so}(4,1)$ representation, one gets in this way the Dirac (spin 1/2) or the DKP (spin 0 or 1) propagator (and presumably the propagator for higher spins as well).

To complete the derivation we need yet another property of the correlation function (\ref{correlation})
derived in the Appendix \ref{sec:correlation},
\begin{equation}
\left<i\left|\exp\left(ip_{\mu}\int_{0}^{L}\hat{\upsilon}^{\mu}\left(t\right)dt\right)\right|j\right>=\left<i\left|\exp\left(iLp\cdot\hat{\upsilon}\right)\right|j\right>\,.\label{eq:correlationOperators}
\end{equation}
Now we can integrate (\ref{eq:correlationOperators}) over $L$ to
find the momentum space propagator
\begin{equation}
G_{ij}(p)=\left<i\left|\frac{i}{p\cdot\hat{\upsilon}-m-i\epsilon}\right|j\right>.
\label{kSpinPropagatorP}
\end{equation}

\section{The propagator for different spins }
\label{sec:DKPpropagator}

Depending on the specific choice of representation for the $\so$(4,1)
generators, expression (\ref{kSpinPropagatorP}) gives the propagator
for different spin values in the first order formalism. As shown in
Appendix \ref{sec:OrbitsIrreps}, the spinning term $\exp\left(i\int\omega\right)$,
upon appropriate choice of coadjoint orbits, decomposes the path-integral
into matrix elements between states belonging to the finite dimensional
representations of SO(4,1) labeled by the highest weights of the irreducible
representations of the maximally compact subgroup SO(4)$\simeq$SU(2)$\otimes$SU(2),
parametrized by a set of ordered integer or half-integer numbers\footnote{The characterization of these finite dimensional representations is
carried out in Appendix \ref{sec:OrbitsIrreps}. The classification
of irreducible unitary (infinite dimensional) representations of SO(4,1)
induced from the maximal compact subgroup was accomplished in~\cite{DixmierDeSitter} following the method developed in \cite{ThomasDeSitter,NewtonDeSitter}, exploiting the relation between
representations of the group and of its Lie algebra. The analogous
characterization for the Euclidean case, leading to (finite dimensional)
irreps of SO(5), is carried out in~\cite{KemmerSO5}.}
\begin{equation}
\left(p,q\right):\qquad p\geq q\geq0.
\end{equation}

Following the argument worked out in~\cite{Fischbach:1974cy} for the Euclidean
case we can define the algebras ${\cal B}^{(k)}$ arising from the
$\so(4,1)$ matrix representations $\pi_{p,q}\left(\hat{\upsilon}_{\mu}\right)$
defined in App. \ref{sec:OrbitsIrreps}, satisfying the commutation
relations, following from (\ref{so41commutators}).
\begin{equation}
\left[\left[\hat{\upsilon}_{\mu},\hat{\upsilon}_{\nu}\right],\hat{\upsilon}_{\rho}\right]=\hat{\upsilon}_{\mu}\eta_{\nu\rho}-\hat{\upsilon}_{\nu}\eta_{\mu\rho}.\label{commutationAlpha}
\end{equation}
and, for $k\geq p$, the equation (following from (\ref{CharactEq}))
\begin{equation}
\prod_{m=-k}^{k}\left(\hat{\upsilon}_{\mu}-\left(\delta_{\mu0}+i\sum_{i}\delta_{\mu i}\right)m\id\right)=0.\label{charactEq}
\end{equation}
Different values of $k$ then correspond to different spin sectors.

For $k=1/2$ one has a four dimensional $\pi(\frac{1}{2},\frac{1}{2})$
representation (see App. \ref{sec:OrbitsIrreps}) of ${\cal B}^{(\frac{1}{2})}$
corresponding to the Dirac algebra. Indeed Eq. (\ref{charactEq})
becomes $\hat{\upsilon}_{\mu}^{2}=-\frac{1}{4}\eta_{\mu\mu}$, and defining
$\gamma_{\mu}=2\hat{\upsilon}_{\mu}$, we find from (\ref{commutationAlpha})
that
\begin{equation}
\gamma_{\mu}\gamma_{\nu}+\gamma_{\nu}\gamma_{\mu}=2\eta_{\mu\nu}.
\end{equation}
Plugging this to (\ref{kSpinPropagatorP}) we get the spin$\frac{1}{2}$
propagator
\begin{equation}
G\left(p\right)\propto\frac{1}{p_{\mu}\gamma^{\mu}-m}\left(=\frac{p_{\mu}\gamma^{\mu}+m}{p^{2}-m^{2}}\right).
\label{DiracPropagator}
\end{equation}
It appears that the spin-$\frac{1}{2}$ propagator has its usual form
expressed in terms of $p$ momenta.

For $k=1$ the matrices $\beta_{\mu}$ (no summation)
satisfy the relations that define the DKP (Duffin-Kemmer-Petiau) algebra~\cite{Duffin:1938zz,kemmerDKP,Petiau}
\begin{equation}
\beta_{\mu}\beta_{\rho}\beta_{\nu}+\beta_{\nu}\beta_{\rho}\beta_{\mu}=\beta_{\mu}\eta_{\nu\rho}+\beta_{\nu}\eta_{\mu\rho}.\label{eq:DKPbeta}
\end{equation}
The derivation of Eq. (\ref{eq:DKPbeta}) is carried out in App. \ref{sec:DKP}.
In this case one has three irreducible representations (see App. \ref{sec:OrbitsIrreps}),
the trivial one-dimensional $\pi\left(0,0\right)$, the five dimensional
 $\pi\left(1,0\right)$, and the ten dimensional
$\pi\left(1,1\right)$. Several results (see for instance \cite{LunardiGauge} and \cite{umezawa}) have been
obtained showing that for these two latter irreducible representations the DKP field equations
reduce respectively to the equations of motion for a spin-0 scalar
field (the Klein-Gordon equation) and for a spin-1 vector field (the
Proca equations). However to our knowledge the reduction of the propagator
to the standard expressions for the spin-0 and spin-1 fields have
not been treated thoroughly, and we devote next section to this task.

\subsection{The propagator for spin-0 and spin-1}

Let us start noticing that it follows from (\ref{kSpinPropagatorP}) that the propagator
in momentum space is (apart from the term $i\epsilon$) the inverse
of the matrix $\p-m$, where we denote $\p=p_{\mu}\beta^{\mu}$, i.e.
\begin{equation}
G\left(p\right)=\left(\p-m\id\right)^{-1}.
\end{equation}
Using the properties (\ref{eq:DKPbeta}) of the DKP matrices, one
can prove that (see for instance \cite{DKPscattering})
\begin{equation}
G\left(p\right)=\frac{1}{m}\left(\frac{\p\left(\p+m\right)}{p^{2}-m^{2}}-\id\right).\label{DKPpropagator}
\end{equation}
Indeed, from (\ref{eq:DKPbeta}),
\begin{equation}
\p^{3}=p_{\mu}p_{\rho}p_{\nu}\beta^{\mu}\beta^{\rho}\beta^{\nu}=p_{\mu}p_{\rho}p_{\nu}\beta^{\mu}\eta^{\rho\nu}=\p p^{2}.
\end{equation}
Then
\begin{equation}
\p\left(\p+m\id\right)\left(\p-m\id\right)=\left(\p^{3}-m^{2}\p\right)=\p\left(p^{2}-m^{2}\right),
\end{equation}
and it follows
\begin{equation}
G\left(p\right)\left(\p-m\id\right)=\frac{1}{m}\left(\p-\left(\p-m\id\right)\right)=\id.
\end{equation}

We consider first the 5 dimensional representation $\pi\left(1,0\right)$
describing the spin-0 sector. The field equations for spin-0 are obtained
with the help of a projection operator~\cite{LunardiGauge}
\begin{equation}
{\cal P}=-\beta_{0}^{2}\beta_{1}^{2}\beta_{2}^{2}\beta_{3}^{2},
\end{equation}
so that the field
\begin{equation}
\psi=\left(\begin{array}{c}
\psi_{0}\\
\psi_{1}\\
\psi_{2}\\
\psi_{3}\\
\psi_{4}
\end{array}\right)
\end{equation}
decomposes into the vector field ${\cal V}^{\mu}={\cal P}\beta^{\mu}\psi$
and the scalar one $\Phi={\cal P}\psi$. Indeed one can show that
$\Phi$ and ${\cal V}^{\mu}$ transform, respectively, as a (pseudo)-scalar
and a (pseudo)-vector under Lorentz transformations, where infinitesimal
transformations are generators by $S_{\mu\nu}=\left[\beta_{\mu},\beta_{\nu}\right]$ as ($\omega^{\mu\nu}=-\omega^{\mu\nu}$)
\begin{equation}
U\simeq 1+\frac{1}{2}\omega^{\mu\nu}S_{\mu\nu}.
\label{LorentzDKP}
\end{equation}
Moreover, one can show that upon imposing the DKP equation for the
free field $\psi$ the components of $\psi$ are not independent,
and that one can define (see App. \ref{sec:DKP}) a specific
representation of the $\beta^{\mu}$ such that $\psi_{4}=\phi$ and
$\psi_{\mu}=\partial_{\mu}\phi$, making explicit the fact that $\psi$
describes in this case the scalar $\phi$ and its derivatives
$\partial_{\mu}\phi$.

We can obtain the propagator for the scalar field ${\cal S}\left(p\right)$
by projecting the propagator (\ref{DKPpropagator}) on the scalar
field sector with ${\cal P}$,
\begin{equation}
{\cal S}\left(p\right)\equiv\frac{1}{m}{\cal P}G\left(p\right){\cal P}^{\dagger},
\end{equation}
so that ${\cal S}\left(p\right)$ is defined by the matrix element in (the
mass factor is for dimensional reasons)
\begin{equation}
\bar{\Phi}{\cal P}G\left(p\right){\cal P}^{\dagger}\Phi=m\bar{\Phi}{\cal S}\left(p\right)\Phi.\label{scalarPropagator1}
\end{equation}
As discussed in App. \ref{sec:OrbitsIrreps} since we are in Lorentzian
metric the $\beta_{0}$ and $\beta_{j}$ matrices must have opposite
hermiticity, and in particular in our notations we have that $\beta_{0}$
is Hermitian and $\beta_{j}$ anti-Hermitian: $\beta_{0}^{\dagger}=\beta_{0},\qquad\beta_{j}^{\dagger}=-\beta_{j}$.

In DKP theory the adjoint field is given by $\bar{\psi}=\psi^{\dagger}\eta_{0}$,
where $\eta_{\mu}$ is the operator
\begin{equation}
\eta_{\mu}=2\beta_{\mu}^{2}-\eta_{\mu\mu},\label{etaDKP}
\end{equation}
such that $\beta_{\mu}^{\dagger}=\eta_{0}\beta_{\mu}\eta_{0}$, and
$\bar{\Phi}=\psi^{\dagger}\eta_{0}{\cal P}^{\dagger}$. Noticing also
that from the defining properties of the $\beta$ matrices (\ref{eq:DKPbeta}),
setting $\mu=\nu$ in (\ref{eq:DKPbeta}), follow the relations
\begin{equation}
\beta_{\mu}\beta_{\nu}\beta_{\mu}=\beta_{\mu}\eta_{\mu\nu},\label{betamnm}
\end{equation}
one finds that
\begin{equation}
{\cal P}\p{\cal P}^{\dagger}=p^{\mu}{\cal P}\beta_{\mu}{\cal P}^{\dagger}=0.\label{scalarPropb}
\end{equation}
Using again (\ref{eq:DKPbeta}) (setting $\nu=\rho\neq\mu$ in (\ref{eq:DKPbeta})
) we find the relations
\begin{equation}
\beta_{\mu}\beta_{\nu}^{2}+\beta_{\nu}^{2}\beta_{\mu}=\beta_{\mu}\eta_{\nu\nu}\qquad\mu\neq\nu,\label{betamn2}
\end{equation}
and multiplying last relation by $\beta_{\mu}$ from the left and
from the right we find
\begin{equation}
\beta_{\mu}^{2}\beta_{\nu}^{2}=\beta_{\nu}^{2}\beta_{\mu}^{2}.\label{betam2n2}
\end{equation}
From last relation, the Hermiticity of $\beta$'s and (\ref{betamnm})
we find also that
\begin{equation}
{\cal P}^{\dagger}={\cal P},\qquad{\cal P}^{2}={\cal P},
\end{equation}
while using (\ref{betam2n2}) and (\ref{betamnm}) it follows
\begin{equation}
{\cal P}\beta^{\mu}\beta^{\nu}={\cal P}\eta^{\mu\nu}.
\end{equation}
From last relations we find
\begin{equation}
{\cal P}\p\p{\cal P}^{\dagger}=p^{2}{\cal P}{\cal P}^{\dagger}=p^{2}.\label{scalarPropbb}
\end{equation}
Plugging (\ref{scalarPropb}) and (\ref{scalarPropbb}) together with
(\ref{DKPpropagator}) in (\ref{scalarPropagator1}) we finally obtain
\begin{equation}
\frac{1}{m}\bar{\Phi}{\cal P}G\left(p\right){\cal P}^{\dagger}\Phi=\frac{1}{m^{2}}\bar{\Phi}{\cal P}\left(\frac{\p\left(\p+m\right)}{p^{2}-m^{2}}-\id\right){\cal P}^{\dagger}\Phi=\bar{\Phi}\frac{1}{p^{2}-m^{2}}\Phi,
\end{equation}
so that
\begin{equation}
{\cal S}\left(p\right)=\frac{1}{p^{2}-m^{2}}.
\end{equation}
The DKP propagator, projected on the scalar field sector, has the
standard form.

We can repeat a similar procedure to derive the propagator for the
spin-1 representation. In this case the projection operators are
\begin{equation}
{\cal R}_{\mu}=\beta_{1}^{2}\beta_{2}^{2}\beta_{3}^{2}\left(\beta_{\mu}\beta_{0}-\eta_{\mu0}\right),
\end{equation}
where now the $\beta$ matrices are to be taken in the 10 dimensional
irreducible representation (we give an explicit realization in App.
\ref{sec:DKP}). The beta matrices maintain the same hermiticity
of the scalar case, and one can show that ${\cal R}_{\mu}\psi$ transforms,
under the infinitesimal Lorentz transformation (\ref{LorentzDKP}),
like a (pseudo)vector while ${\cal R}_{\mu\nu}\psi={\cal R}_{\mu}\beta_{\nu}\psi$
like a (pseudo)tensor. Upon imposing the DKP equation for $\psi$
one can then show that ${\cal R}_{\mu\nu}$ is proportional to the
strength tensor of the vector field $R_{\mu}\psi$ (see for instance \cite{LunardiGauge}
and~\cite{umezawa}). We define then the vector field ${\cal A}_{\mu}$
and its adjoint as
\begin{equation}
{\cal A}_{\mu}={\cal R}_{\mu}\psi,\qquad\bar{{\cal A}}_{\mu}=-\bar{\psi}{\cal R}_{\mu}^{\dagger}=-\psi^{\dagger}\eta_{0}{\cal R}_{\mu}^{\dagger},
\end{equation}
with $\eta_{0}$ given by (\ref{etaDKP}). It is possible to show
that with this definition the fields ${\cal A}_{\mu}$ and $\bar{{\cal A}}_{\mu}$
transform respectively with covariant and contravariant indexes. Thus
we may identify
\begin{equation}
{\cal A}^{\mu}={\cal \bar{A}_{\mu}},
\end{equation}
($\sum_{\mu}\bar{{\cal A}}_{\mu}{\cal A}_{\mu}={\cal A}_{\mu}{\cal A}^{\mu}$
transforms as a (pseudo)scalar). The spin-1 propagator ${\cal S}^{\mu\nu}\left(p\right)$
for the vector field is then obtained by projection
\begin{equation}
{\cal S}_{\mu}^{\ \nu}\left(p\right)\equiv\frac{1}{m}{\cal R}_{\mu}G\left(p\right){\cal R}_{\nu}^{\dagger}
\end{equation}
as the matrix element in
\begin{equation}
\sum_{\mu,\nu}{\cal \bar{A}_{\mu}}{\cal R}_{\mu}G\left(p\right){\cal R}_{\nu}^{\dagger}{\cal A}_{\nu}=m{\cal A_{\mu}}{\cal S}^{\mu\nu}\left(p\right){\cal A}_{\nu}.
\end{equation}
We can use the properties of the $\beta$ matrices (\ref{betamnm}),
(\ref{betamn2}) and (\ref{betam2n2}) to find the following relations
\begin{equation}
\begin{gathered}{\cal R}_{\mu}{\cal R}_{\nu}^{\dagger}=\delta_{\mu}^{\ \nu}{\cal R}_{0},\qquad{\cal R}_{0}{\cal R}_{\mu}={\cal R}_{\mu}\qquad{\cal R}_{\mu}\beta_{\rho}{\cal R}_{\nu}^{\dagger}=0,\\
{\cal R}_{\mu}\beta_{\rho}\beta_{\sigma}=\eta_{\rho\sigma}{\cal R}_{\mu}-\eta_{\mu\sigma}{\cal R}_{\rho}.
\end{gathered}
\end{equation}
Using these relations it follows that
\begin{equation}
\begin{gathered}{\cal R}_{\mu}\p{\cal R}_{\nu}^{\dagger}=0,\\
{\cal R}_{\mu}\p\p{\cal R}_{\nu}^{\dagger}=p^{2}{\cal R}_{\mu}{\cal R}_{\nu}^{\dagger}-p_{\mu}p^{\rho}{\cal R}_{\rho}{\cal R}_{\nu}^{\dagger},
\end{gathered}
\end{equation}
and finally
\begin{equation}
\begin{split}\frac{1}{m}{\cal \bar{A}_{\mu}}{\cal R}_{\mu}G\left(p\right){\cal R}_{\nu}^{\dagger}{\cal A}_{\nu}= & \frac{1}{m^{2}}{\cal \bar{A}_{\mu}}{\cal R}_{\mu}\left(\frac{\p\left(\p+m\right)}{p^{2}-m^{2}}-\id\right){\cal R}_{\nu}^{\dagger}{\cal A}_{\nu}\\
= & \frac{1}{m^{2}\left(p^{2}-m^{2}\right)}{\cal \bar{A}_{\mu}}\left(m^{2}{\cal R}_{\mu}{\cal R}_{\nu}^{\dagger}-p_{\mu}p^{\rho}{\cal R}_{\rho}{\cal R}_{\nu}^{\dagger}\right){\cal A}_{\nu}\\
= & \frac{1}{m^{2}\left(p^{2}-m^{2}\right)}{\cal \bar{A}_{\mu}}\left(m^{2}\delta_{\mu}^{\ \nu}-p_{\mu}p^{\nu}\right){\cal R}_{0}{\cal A}_{\nu}\\
= & {\cal A_{\mu}}\frac{1}{p^{2}-m^{2}}\left(\eta^{\mu\nu}-\frac{p^{\mu}p^{\nu}}{m^{2}}\right){\cal A}_{\nu},
\end{split}
\end{equation}
where we used the above relations together with ${\cal R}_{0}{\cal A}_{\mu}={\cal R}_{0}{\cal R}_{\mu}\psi={\cal R}_{\mu}\psi={\cal A}_{\mu}$.
Thus the spin-1 propagator reduces to the standard propagator in unitary
gauge
\begin{equation}
{\cal S}^{\mu\nu}\left(p\right)=\eta^{\mu\nu}-\frac{p^{\mu}p^{\nu}}{m^{2}},
\label{spin1propagator}
\end{equation}

\section{Conclusions}

In this paper, following the idea of Alekseev and Shatashvili of adding to the first order action the Kirillov presymplectic form, which forces the path integral to select a particular representation of the de Sitter group, we derive the DKP propagator for fields of spin 0 and 1 (as well as the Dirac propagator for spin 1/2) in the path integral formalism. We then show that the obtained DKP propagators are equivalent to the standard ones.

There are several interesting problems that could be addressed in follow-up investigations. First, although it seems pretty obvious that an analogous construction should work for spins higher than 1, it would be illuminating to do it explicitly.

Second, the construction presented here can, presumably, be extended to the case of $\kappa$-deformation (in the sense of $\kappa$-Poincar\'e Hopf symmetries)~\cite{Lukierski:1991pn},~\cite{Lukierski:1992dt},~\cite{Majid:1994cy},~\cite{Kowalski-Glikman:2017ifs},
a scenario that has attracted much interest especially in relation with quantum gravity phenomenology.
In this case momentum space is not the ordinary flat (Minkowskian) momentum space, but it is described as a curved manifold (specifically the group $AN_3$, corresponding to half of de Sitter space, see~\cite{Kowalski-Glikman:2017ifs}), whose scale of curvature $1/\kappa$ is taken to be proportional to the (inverse) Planck energy ($1/E_{pl} \sim 10^{-19} GeV$).
As (four-dimensional) $\kappa$-momentum space can be also described in terms of flat embedding `momentum coordinates' in five dimensions, with some additional constraint enforcing physical momenta to live on the de Sitter hyperboloid, one can think of extending the formalism described in this manuscript, which is not restricted to four-dimensional momentum space, exploiting the use of embedding coordinates.
The construction of a Dirac (spin 1/2) action with $\kappa$-Poincar\'e symmetries has been already addressed in some previous works (see for instance~\cite{Lukierski:1992dt,Nowicki:1992if,Kosinski:1998kw,Agostini:2002yd}).
It would be interesting to compare the spin 1/2 propagator for $\kappa$-momentum space resulting from our approach with previous results.
Moreover, if working, our construction would allow in principle to study higher spin propagators for $\kappa$-momentum space, setting the stage for constructing a higher-spin field theory action based on $\kappa$-deformed symmetries.

\section*{Acknowledgment}

For JKG this work is supported by Polish National Science Centre, projects number 2017/27/B/ST2/01902.

\appendix

\section{The action functional on the orbits}

\label{sec:ActionOrbit}

We here discuss the construction of the action (\ref{actionS}) needed
to implement the spin degrees of freedom in the path-integral. We
refer the reader to the characterization outlined for instance in~\cite{Alekseev:1988ce},
based on the Kirillov symplectic form~\cite{Kirillov}. Consider
a (matrix) Lie group $G$. Let $\mathfrak{g}$ be Lie algebra of $G$
and $\mathfrak{g}^{*}$ its dual Lie algebra: for a basis $\left\{ e_{a}\right\} $
of $\mathfrak{g}$ and $\left\{ \tilde{e}^{a}\right\} $ of $\mathfrak{g}^{*}$,
the duality relations are canonically given by $\left\langle \tilde{e}^{a},e_{b}\right\rangle =\delta_{b}^{a}$.
The coadjoint representation of $G$ is defined by
\begin{equation}
\begin{gathered}\begin{gathered}\left\langle \text{Ad}^{*}\left(g\right)X,u\right\rangle =\left\langle X,\text{Ad}\left(g\right)^{-1}u\right\rangle ,\qquad\text{where}\qquad\text{Ad}\left(g\right)u=gug^{-1},\end{gathered}
\\
\left\langle \text{ad}^{*}\left(u\right)X,v\right\rangle =-\left\langle X,\text{ad}\left(u\right)v\right\rangle ,\qquad\text{where}\qquad\text{ad}\left(u\right)b=\left[u,v\right],
\end{gathered}
\end{equation}
where $g\in G$, $X\in\mathfrak{g}^{*}$, $u,v\in\mathfrak{g}$. Let
us parametrize the orbits by group variables fixing the point $X_{0}$
so that a generic point on the orbit is
\begin{equation}
X\left(g\right)=\text{Ad}^{*}\left(g\right)X_{0}.\label{eq:GroupCoordinate}
\end{equation}
In the basis $\left\{ e_{a}\right\} $ and $\left\{ \tilde{e}^{a}\right\} $
we will write a generic point on the orbit as $X=l_{a}\tilde{e}^{a}$.
Define the action
\begin{equation}
S=\int\omega,\qquad\omega=\left\langle X\left(g\right),Y\left(g\right)\right\rangle \label{eq:action}
\end{equation}
with
\begin{equation}
Y\left(g\right)=dgg^{-1}\in\mathfrak{g}.
\end{equation}
Here $Y=dgg^{-1}$ is the Maurer-Cartan form on the group. It is possible
to show that the following equivalent equations are satisfied
\begin{equation}
dX=\text{ad}^{*}\left(Y\right)X\label{eq:conditiondX}
\end{equation}
\begin{equation}
dl_{a}=f_{ab}^{\ \ c}Y^{b}l_{c},\label{eq:conditiondl}
\end{equation}
where $X=l_{a}\tilde{e}^{a}$, and $Y=Y^{a}e_{a}$, and $f_{ab}^{\ \ c}$
are the structure constant of the Lie algebra\footnote{Here $[\cdot,\cdot]$ denotes obviously the Lie bracket.}
$\mathfrak{g}$ $\left[e_{a},e_{b}\right]=f_{ab}^{\ \ c}e_{c}$. Let
us first show the equivalence of Eq.(\ref{eq:conditiondl}) and Eq.(\ref{eq:conditiondX}):
\begin{equation}
\begin{split}dl_{a}= & \left\langle dX,e_{a}\right\rangle =\left\langle \text{ad}^{*}\left(Y\right)X,e_{a}\right\rangle =\left\langle X,\left[e_{a},Y\right]\right\rangle \\
= & l_{c}Y^{b}\left\langle \tilde{e}^{c},\left[e_{a},e_{b}\right]\right\rangle =l_{c}Y^{b}f_{ab}^{\ \ d}\left\langle \tilde{e}^{c},e_{d}\right\rangle =f_{ab}^{\ \ c}Y^{b}l_{c}.
\end{split}
\end{equation}
Let's now prove Eq.(\ref{eq:conditiondX}):
\begin{equation}
\begin{split}\left\langle dX,e_{a}\right\rangle = & d\left\langle \text{Ad}^{*}\left(g\right)X_{0},e_{a}\right\rangle =\left\langle X_{0},g^{-1}e_{a}dg+dg^{-1}e_{a}g\right\rangle \\
= & \left\langle \text{Ad}^{*}\left(g\right)X_{0},\left[e_{a},dgg^{-1}\right]\right\rangle =-\left\langle X,\text{ad}\left(Y\right)e_{a}\right\rangle =\left\langle \text{ad}^{*}\left(Y\right)X,e_{a}\right\rangle ,
\end{split}
\end{equation}
where we used definition (\ref{eq:GroupCoordinate}) and $d\mathbbm{1}=d\left(g^{-1}g\right)=dg^{-1}g+g^{-1}dg=0$,
from which follows $dg^{-1}=-g^{-1}dgg^{-1}$.

Eq. (\ref{eq:conditiondX}) (or (\ref{eq:conditiondl})) ensures that
the action (\ref{eq:action}) generates on the orbit the canonical
2-form
\begin{equation}
\Omega=-d\omega=\left\langle X,Y\wedge Y\right\rangle ,\label{eq:canonical2form}
\end{equation}
where $d\Omega$ is closed on the orbit. The last equation can be
rewritten explicitly as
\begin{equation}
\begin{split}\Omega=-d\omega= & \left\langle X_{0},g^{-1}dg\wedge g^{-1}dg\right\rangle \\
= & \left\langle X,dgg^{-1}\wedge dgg^{-1}\right\rangle ,
\end{split}
\end{equation}
where the Maurer-Cartan equation $d(gdg^{-1})=-gdg^{-1}\wedge gdg^{-1}$
has been used. One can show that the Poisson brackets of the restriction
of the linear functions on the orbit reproduce the algebraic commutation
relation. Defining the linear functions $u\left(X\right)=\left\langle X,u\right\rangle $,
where on the r.h.s. $u=u^{a}e_{a}$ (so that $u\left(X\right)=u^{a}l_{a}$),
we get
\begin{equation}
\begin{split}\left\{ u\left(X\right),v\left(X\right)\right\} = & \Omega\left(u,v\right)=\left\langle X,\left[Y\left(u\right),Y\left(v\right)\right]\right\rangle \\
= & u^{a}v^{b}\left\langle X,\left[e_{a},e_{b}\right]\right\rangle =f_{ab}^{\ \ c}u^{a}v^{b}l_{c}
\end{split}
\end{equation}
where we used that the Maurer-Cartan form evaluated on an element
of the basis of the Lie algebra gives $Y\left(e_{a}\right)=e_{a}$.
It follows in particular, for $u^{a}=\delta_{b}^{a}$, that on the
orbit,
\begin{equation}
\left\{ l_{a},l_{b}\right\} =f_{ab}^{\ \ c}l_{c}.\label{eq:PoissonBrackets}
\end{equation}

\section{Coadjoint orbits and irreps for SO(4,1)}
\label{sec:OrbitsIrreps}

\subsection{Integral orbits for SO(3,2)}

In the spirit of geometric quantization~(see for instance~\cite{Woodhouse:1992de,HallQM}
the orbit
method~\cite{Kirillov} can be used to ``quantize'' the values
of some parameters labeling the orbits of the action of the group
on its dual Lie algebra. This mechanism can be realized~\cite{Alekseev:1988tj,Alekseev:1988vx}  by the requirement for the action exponential $\exp\left(iS\left(\ell\right)\right)$
to be single valued, so that the path-integral is well defined. Indeed
the 1-form $\omega$ is singular, and the action $S=\int\omega$ is
multivalued, and the requirement of uniqueness of the expression $\exp\left(iS\left(\ell\right)\right)$
over closed path leads to integral orbits.

In our specific case, starting from the $\so(3,2)$ algebra~(\ref{cl(3,1)algebra}), so that a generic element can be parametrized as $u = \tilde{\upsilon}^\mu \Upsilon_\mu + \tfrac{1}{2}\tilde{j}^{\mu\nu} J_{\mu\nu}$ ($\tilde{j}^{\nu\mu}=-\tilde{j}^{\mu\nu}$), we can fix the orbits considering the action of the Lorentz subgroup SO(3,1) generated by $J_{\mu\nu}$.
Rewriting the generators as $R_{i}=-\frac{1}{2}\epsilon_{ijk}J_{jk}$
and $P_{i}=\Upsilon_i$, we can rewrite the $\so(3,1)$ subalgebra as
\begin{equation}
\left[R_{i},R_{j}\right]=\epsilon_{ijk}R_{k},\qquad\left[R_{i},P_{j}\right]=\epsilon_{ijk}P_{k},\qquad\left[P_{i},P_{j}\right]=-\epsilon_{ijk}R_{k},
\end{equation}
so that an element of the $\so(3,1)$ subalgebra is $u_{\so(3,1)} = r^i R_i + p^i P_i $, with $r^i = - \epsilon^i_{jk}\tilde{j}^{jk}$, $p^i = \tilde{v}^i$.
Reparametrizing~(see~\cite{WeinbergBook} Ch. 5.6) an element of $\so(3,1)$ as $u_{\so(3,1)}=a^{i}A_{i}+b^{i}B_{i}$,
with $A_{i}=\frac{1}{2}(R_{i}+ i P_{i})$, $B_{i}=\frac{1}{2}(R_{i} - i P_{i})$, so that $a^i = r^i + i p^i$, $b^i = r^i - i p^i$,
the algebra splits into a direct sum of 2 mutually commuting complex (conjugate) $\su$(2):
$\so(3,1) \approx \su(2)_{\C}\oplus \overline{\su(2)}_{\C} \approx \sl(2,\C)\oplus \overline{\sl(2,\C)}$. Thus we have reduced the problem
to fixing the orbits of the two $Sl(2,\C)$ subgroups of $SO(3,1)$.
Finally, we notice that each of the two $Sl(2,\C)$ admits $SU(2)$ as (maximal) compact subgroup, and we can use it to fix the orbits for each of the two copies.

Representing
the SU(2) generators in terms of Pauli matrices, $A_{i},B_{i}\equiv-\frac{i}{2}\sigma_{i}$,
\begin{equation}
\sigma_{1}=\left(\begin{array}{cc}
0 & 1\\
1 & 0
\end{array}\right),\qquad\sigma_{2}=\left(\begin{array}{cc}
0 & -i\\
i & 0
\end{array}\right),\qquad\sigma_{3}=\left(\begin{array}{cc}
1 & 0\\
0 & -1
\end{array}\right),
\end{equation}
for each SU(2) copy we can parametrize an element of the group by
Euler angles as
\begin{equation}
g_{\text{SU}(2)}=\exp\left(-\frac{i}{2}\sigma_{3}\chi\right)\exp\left(-\frac{i}{2}\sigma_{2}\theta\right)\exp\left(-\frac{i}{2}\sigma_{3}\phi\right),
\end{equation}
with
\begin{equation}
\chi\in[0,2\pi),\qquad\theta\in[0,\pi],\qquad\phi\in[0,2\pi).
\end{equation}

On each copy the Maurer-Cartan connection $Y_{\text{SU}(2)}=g_{\text{SU}(2)}^{-1}dg_{\text{SU}(2)}$
can be evaluated to
\begin{equation}
Y_{\text{SU}(2)}^{i}=\left(\begin{array}{c}
-\sin\chi d\theta+\cos\phi\sin\theta d\phi\\
\cos\chi d\theta+\sin\phi\sin\theta d\phi\\
d\chi+\cos\theta d\phi
\end{array}\right),
\end{equation}
where $Y=Y^{i}A_{i}$ or $Y=Y^{i}B_{i}$.
The orbits can be chosen
fixing the value of the coordinates in $\g_{\su(2)}^{*}$ $\tilde{a}=\tilde{a}_{i}\tilde{A}^{i}$,
$\tilde{b}=\tilde{b}_{i}\tilde{A}^{i}$ along the (real) $z$ direction ($\braket{\tilde{A}^{i},A_{j}}=\delta_{j}^{i}$,
$\braket{\tilde{B}^{i},B_{j}}=\delta_{j}^{i}$), $Re(\tilde{a}_{i})=\left(0,0,m\right)$,
$Re(\tilde{b}_{i})=\left(0,0,n\right)$, and we thus find respectively
the action ($s=m$ or $n$)
\begin{equation}
\int\omega_{SU(2)}=\gamma\int d\phi+s\int\cos\theta d\phi
\end{equation}
where we renamed the azimuthal angle $\chi=\gamma'\phi$ for some
constant $\gamma'$ and $\gamma=\gamma's$. The action is multivalued
as it counts the windings around the axis passing through the poles
$\theta=0,\pi$ of the sphere, where the 1-form $\cos\theta d\phi$
is singular. For infinitesimal closed contours around the poles $\theta=0,\pi$,
the action gives the value $2\pi\left(\gamma\pm s\right)$, so that,
if $\gamma\pm s$ is an integer, the action $\exp\left(i\int\omega\right)$
does not contribute to the path-integral, which is then well-defined.
We can choose $\gamma=0$ for $s$ integer and $\gamma=\frac{1}{2}$
for $s$ semi-integer. Thus the condition for single-valuedness of
$\exp\left(i\int\omega\right)$ translates into the condition of ``quantization''
of the values of $s=\left(m,n\right)$, which take only integer or
semi-integer values $(\ell_{1},\ell_{2})$.

\subsection{Discrete series and finite dimensional representations of SO(4,1)}

With this choice of orbits, after quantization (see Eq.~(\ref{so32commutators})), the elements $l_{a}\equiv\left(\upsilon_{\mu},j_{\mu\nu}\right) \rightarrow (\hat{\upsilon}_{\mu},\hat{j}_{\mu\nu}) $
of the dual algebra belong to one of the irreducible unitary representations of $\so(3,2)$ induced by the (real structure) decomposition $\su(2) \oplus \overline{\su(2)}$. Finally, after the substitution~(\ref{mapSigmaS}) ($\hat{S}_{\mu\nu}=i \hat{j}_{\mu\nu}$) the (matrix) operators $\hat{v}_\mu$ and $S_{\mu\nu}$ belong to one of the irreducible unitary representations of $\so(4,1)$, seen as a Lie algebra with real structure constants, Eq.~(\ref{so41commutators}),
induced by the maximal compact subgroup $\text{SO}(4) \approx \text{SU}(2)\oplus \text{SU}(2)$. The classification
of such representations has been carried out in~\cite{DixmierDeSitter} perfecting an approach developed
previously in~\cite{ThomasDeSitter} and~\cite{NewtonDeSitter} relating
the group representation to the representation of its Lie algebra
generators. In particular it is shown that the discrete series representation
$\pi_{\tilde{p},\tilde{q}}$ of SO(4,1) can be obtained in this way
(see also~\cite{EvansDeSitter}). The discrete series $\pi_{\tilde{p},\tilde{q}}$
is labeled by two integers or semi-integers values $(\tilde{p},\tilde{q})$
related to the SO(4)$\simeq$SU(2)$\otimes$SU(2) labels by $\tilde{p}=\inf_{(\ell_{1},\ell_{2})\in\Gamma}(\ell_{1}+\ell_{2})$,
$\tilde{q}=\inf_{(\ell_{1},\ell_{2})\in\Gamma}(\ell_{1}-\ell_{2})$,
where $\Gamma$ is the set of values $(\ell_{1},\ell_{2})$ which
occur in the reduction $\pi_{\tilde{p},\tilde{q}}|_{\text{SO}(4)}$
of the representation to SO(4), and $\tilde{p}=\frac{1}{2},1,\frac{3}{2},2,\dots$,
$\tilde{q}=\tilde{p},\tilde{p}-1,\dots,1\ \text{or\ \ensuremath{\frac{1}{2}}}$.
The Hilbert space of the representation is the infinite\footnote{Since SO(4,1) is non-compact irreducible unitary representations are
infinite dimensional.} direct sum ${\cal H}=\oplus_{(\ell_{1},\ell_{2})\in\Gamma}{\cal H}_{\ell_{1},\ell_{2}}$
of the subspaces ${\cal H}_{\ell_{1},\ell_{2}}$ invariant under $\pi_{\tilde{p},\tilde{q}}\left(\text{SO}(4)\right)$.
Each representation $\pi_{\tilde{p},\tilde{q}}$ is characterized
by a different value of the two Casimir invariants of the $\so$(4,1)
Lie algebra corresponding to the ones in~(\ref{casimir}) of $\so(3,2)$:
\begin{equation}
\begin{gathered}\hat{{\cal C}}_{1}=\hat{\upsilon}^{\mu}\hat{\upsilon}_{\mu}-\frac{1}{2}\hat{S}_{\mu\nu}\hat{S}^{\mu\nu},\\
\hat{{\cal C}}_{2}=\hat{W}_{0}^{2}-\hat{W}\cdot \hat{W}-\frac{1}{4}\left(\epsilon_{ijk}\hat{S}_{jk}\hat{S}_{i0}\right)^{2},
\end{gathered}
\label{casimirSO41}
\end{equation}
where $\hat{W}_{0}=\frac{1}{2}\epsilon_{ijk}\hat{S}_{jk}\hat{\upsilon}_{i}$, $\hat{W}=-\frac{1}{2}\epsilon_{ijk}\hat{S}_{jk}\hat{\upsilon}_{0}+\epsilon_{ijk}\hat{S}_{j0}\hat{\upsilon}_{k}$.
Indeed, for a given $\pi_{\tilde{p},\tilde{q}}$,
$\hat{{\cal C}}_{1}$ and $\hat{{\cal C}}_{2}$ are scalar operators taking the
values~\cite{DixmierDeSitter}
\begin{equation}
\begin{gathered}\pi_{\tilde{p},\tilde{q}}(\hat{{\cal C}}_{1})=(\tilde{p})(\tilde{p}+1)-2+(\tilde{q})(\tilde{q}-1),\\
\pi_{\tilde{p},\tilde{q}}(\hat{{\cal C}}_{2})=(\tilde{p})(\tilde{p}+1)(\tilde{q})(\tilde{q}-1).
\end{gathered}
\end{equation}
Thus, the orbit quantization method exhibited in the previous section,
singling out integer or semi-integer values $(\ell_{1},\ell_{2})$,
selects the values of the invariant polynomials ${\cal C}_{1}$ and
${\cal C}_{2}$ of $l_{a}\equiv(\upsilon_{\mu},j_{\mu\nu})$, defining
the orbits, which occur in the irreducible representations of SO(4,1),
and in particular in the discrete series.

\begin{figure}[h!]
\includegraphics{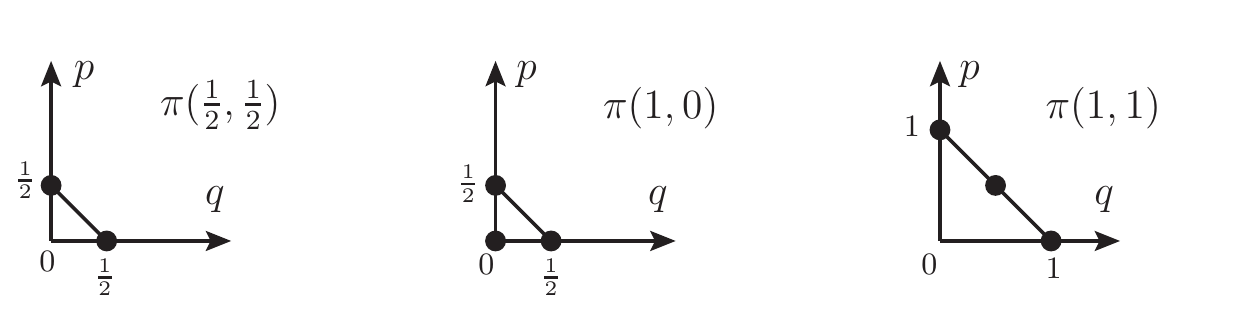}
\caption{\label{fig:irrepsLow}The lowest order $\pi\left(p,q\right)$ representations
of SO(4,1) in terms of SO(4)$\simeq$SU(2)$\otimes$SU(2) labels.
The dots indicate the $(\ell_{1},\ell_{2})$ values within each representation.}
\end{figure}

Apart from the irreducible unitary (infinite dimensional) representations,
one can obtain finite dimensional, non-unitary, representations that
can be understood as a ``Wick-rotation'' of the irreducible unitary
representations of SO(5) discussed for instance in~\cite{KemmerSO5}.
In this case the generators can be represented through finite dimensional
(mixed Hermitian and anti-Hermitian) matrices, and we will use this
representations to construct the projection operators for the different
spin sectors of the propagator. Denoting such finite dimensional representations
with $\pi_{p,q}$, they are labeled now by $p=\max(\ell_{1}+\ell_{2})$
and $q=\max(\ell_{1}-\ell_{2})$, whose ranges are such that $p$
and $q$ are all integers or all semi-integers and $p\geq q\geq0$,
while, for given $\left(p,q\right)$, $\ell_{1}+\ell_{2}$ and $\ell_{1}-\ell_{2}$
range respectively from $q$ to $p$ and from $-q$ to $q$ by steps
of 1 (thus $\ell_{1},\ell_{2}$ range from 0 to $\frac{1}{2}\left(p+q\right)$
by steps of $\frac{1}{2}$). The $\pi_{p,q}$ are are characterized
by the values of the Casimir operators
\begin{equation}
\begin{gathered}\pi_{p,q}({\cal \hat{C}}_{1})=p\left(p+3\right)+q\left(q+1\right),\\
\pi_{p,q}({\cal \hat{C}}_{2})=\left(p+1\right)\left(p+2\right)q\left(q+1\right),
\end{gathered}
\end{equation}
and their dimension is given by the formula
\begin{equation}
d\left(p,q\right)=\frac{1}{6}\left(2q+1\right)\left(2p+3\right)\left(p+q+2\right)\left(p-q+1\right).\label{dimensionpq}
\end{equation}
The range of the highest weights $(\ell_{1},\ell_{2})$ for the lowest
order representations $\pi_{p,q}$ is depicted in Fig. \ref{fig:irrepsLow}
in terms of the allowed $(\ell_{1},\ell_{2})$ values. From the formula
(\ref{dimensionpq}) we find that they have respectively the dimensions
$d\left(\frac{1}{2},\frac{1}{2}\right)=4$, $d\left(1,0\right)=5$
and $d\left(1,1\right)=10$, reflecting the decomposition in terms
of their restriction to SO(4): $\pi\left(\frac{1}{2},\frac{1}{2}\right)\rightarrow D(\frac{1}{2},0)\oplus D(0,\frac{1}{2})$,
$\pi\left(1,0\right)\rightarrow D(\frac{1}{2},0)\oplus D(0,\frac{1}{2})\oplus D(0,0)$,
$\pi\left(1,1\right)\rightarrow D(1,0)\oplus D(0,1)\oplus D(\frac{1}{2},\frac{1}{2})$.

The matrices that form the $\pi_{p,q}$ representation can be considered
to be obtained by ``Wick rotation'' from the SO(5) matrices derived
in~\cite{KemmerSO5}. To better characterize this definition consider
first the defining 5 dimensional representation of SO(4,1) matrices.
They are the matrices preserving the bilinear form given by the five
dimensional Lorentzian metric $\eta^{(5)}\equiv(+,-,-,-,-)$, so that
the generators are matrices ${\cal M}_{AB}$ such that ${\cal M}_{AB}^{T}=-\eta^{(5)}{\cal M}_{AB}\eta^{(5)}$,
and are related with the $\so$(5) (skew-symmetric) matrices $\tilde{{\cal M}}_{AB}$
by $\tilde{{\cal M}}_{AB}=-\eta_{5}{\cal M}_{AB}$. We can set then
$({\cal M}_{AB})_{KL}=\eta_{AK}^{(5)}\delta_{BL}-\eta_{BK}^{(5)}\delta_{AL}$
and $(\tilde{{\cal M}}_{AB})_{KL}=-(\delta_{AK}\delta_{BL}-\delta_{BK}\delta_{AL})$,
and the $\so(4,1)$ commutation rules $\left[{\cal M}_{AB},{\cal M}_{CD}\right]=\eta_{AD}{\cal M}_{BC}+\eta_{BC}{\cal M}_{AD}-\eta_{AC}{\cal M}_{BD}-\eta_{BD}{\cal M}_{AC}$,
as well as the $\so(5)$ ones for $\tilde{{\cal M}}_{AB}$, are satisfied.
On passing from $\so(5)$ to $\so(4,1)$ matrices one can easily see
that the matrices ${\cal M}_{0A}$ become symmetric (and Hermitian)
while, the remaining ones stay skew-symmetric (and anti-Hermitian).
In particular, if $\tilde{\lambda}$ are the (imaginary) eigenvalues
of the $\tilde{{\cal M}}_{AB}$, from the spectral theorem, the effect
of the map on the eigenvalues is $\tilde{\lambda}\rightarrow\lambda=i\tilde{\lambda}$
for the matrices ${\cal M}_{0A}$, while for the remaining ${\cal M}_{iA}$
($i=1,2,3$; $A\neq0$) they stay the same. With this in mind we can
generalize this definition of Wick-rotation to representations of
any dimensions: one can always choose the matrix generators ${\cal M}_{0A}$
to be Hermitian and the ${\cal M}_{iA}$ ($A\neq0$) anti-Hermitian,
and, when going from SO(5) to SO(4,1) matrix generators, the eigenvalues
of the ${\cal M}_{0A}\left(\equiv\{\pi_{p,q}(\hat{\upsilon}_{0}),\pi_{p,q}(\hat{S}_{0i})\}\right)$
components acquire an extra $i$ factor\footnote{One can check explicitly that this is indeed true for the examples
in this manuscript.}.

We can then use almost the same argument of~\cite{Fischbach:1974cy} to
establish the relation between the matrices ${\cal M}_{\mu4}=\pi_{p,q}(\hat{\upsilon}_{\mu})$
and their characteristic polynomial. We observe first that each SO(5)
matrix ${\cal \tilde{M}}_{AB}$ is unitary equivalent to a member
of a triplet of generators forming angular momentum algebra. It thus
follows, considering also the discussion above, that ${\cal \tilde{M}}_{AB}$
is equivalent to a direct sum of diagonal block matrices, each one
with elements $im$ with $m=j,j-1,\qquad,-j$ for $p\geq j\geq0$.
It follows from the Cailey-Hamilton theorem applied to the matrices
${\cal \tilde{M}}_{\mu4}$, and the above considerations, that in
the irrep $\pi_{p,q}$ the generators $\hat{\upsilon}_{\mu}$ satisfy the
minimal characteristic equation
\begin{equation}
\prod_{m=-p}^{p}\left(\hat{\upsilon}_{\mu}-(\delta_{\mu0}+i\sum_{i}\delta_{\mu i})m\id\right)=0.\label{CharactEq}
\end{equation}

\section{Some properties of the correlation function}

\label{sec:correlation}

\subsection{Ward identities}

Under the transformation (\ref{variationl}) the action (\ref{eq:action})
$S=\int\omega\left(l\right)$ varies as
\begin{equation}
\begin{split}\delta S= & \int\left\langle \delta X,Y\right\rangle =\int\left\langle f_{ab}^{\ \ c}\xi^{b}l_{c}\tilde{e}^{a},Y^{d}e_{d}\right\rangle =\int f_{ab}^{\ \ c}\xi^{b}l_{c}Y^{a}\\
= & -\int\xi^{a}f_{ab}^{\ \ c}Y^{b}l_{c}=-\int\xi^{a}dl_{a}=-\int\xi^{a}\partial_{t}l_{a}dt,
\end{split}
\label{eq:changeAction}
\end{equation}
where we used Eq. (\ref{eq:conditiondl}). Notice also that the variation
of $l_{a}$ (\ref{variationl}) is generated by the coadjoint action
of the vector $\xi\in\mathfrak{g}$:
\begin{equation}
\left\langle \text{ad}^{*}\left(\xi\right)X,e_{a}\right\rangle =\xi^{b}\left\langle X,\left[e_{a},e_{b}\right]\right\rangle =f_{ab}^{\ \ c}\xi^{b}\left\langle X,e_{c}\right\rangle =f_{ab}^{\ \ c}\xi^{b}l_{c}=\delta l_{a}.\label{eq:transformationcoadjoint}
\end{equation}
We exploit now the invariance of the functional integral
\begin{equation}
\delta\int{\cal D}\left(l\right)\prod_{i}e^{iS\left(l\right)}l_{a_{1}}\left(t_{1}\right)\cdots l_{a_{n}}\left(t_{n}\right)=0\label{eq:InvarianceFunctional}
\end{equation}
under this transformation (under the coadjoint action (\ref{eq:transformationcoadjoint})
the measure is invariant). The term $e^{iS\left(l_{a}\right)}$ changes
as in (\ref{eq:changeAction}), i.e.
\begin{equation}
\delta e^{iS}=-ie^{iS}\int\xi^{a}\partial_{t}l_{a}\left(t\right)dt,
\end{equation}
while the terms $l^{a_{i}}\left(t_{i}\right)$ change as
\begin{equation}
\delta l_{a_{i}}\left(t_{i}\right)=f_{a_{i}b}^{\ \ c}\xi^{b}l_{c}\left(t_{i}\right)=\int f_{a_{i}b}^{\ \ c}\xi^{b}l_{c}\left(t\right)\delta\left(t-t_{i}\right)dt.
\end{equation}
Then we get from (\ref{eq:InvarianceFunctional})
\begin{equation}
\int D\left(l\right)e^{iS}\int dt\left(-i\xi^{a}\partial_{t}l_{a}\left(t\right)l_{a_{1}}\left(t_{1}\right)\cdots l_{a_{n}}\left(t_{n}\right)+\sum_{k}f_{a_{k}b}^{\ \ c}\xi^{b}\delta\left(t-t_{k}\right)l_{c}\left(t\right)l_{a_{1}}\left(t_{1}\right)\cdots\bar{l}_{a_{k}}\left(t_{k}\right)\cdots l_{a_{n}}\left(t_{n}\right)\right)=0,
\end{equation}
where $\bar{l}_{a_{1}}\left(t_{k}\right)$ means that this term is
missing from the product. From these relations we obtain the Ward
identities (\ref{ward}).

\subsection{n-point correlation function}

One can show that the n-point correlation function is given by the
time ordered product of the field operators as
\begin{equation}
\braket{i|\hat{l}_{a_{1}}\left(t_{1}\right)\cdots \hat{l}_{a_{n}}\left(t_{n}\right)|j}=\sum_{\left\{ i_{1}\cdots i_{n}\right\} }\prod_{k=1}^{n-1}\theta\left(t_{i_{k}}-t_{i_{k+1}}\right)\braket{i|\hat{l}_{a_{i_{1}}}\cdots \hat{l}_{a_{i_{n}}}|j},\label{eq:wardOperators}
\end{equation}
where on the r.h.s. $\hat{l}_{a}$ are matrix operators for a specific representation
of $\mathfrak{so}\left(4,1\right)$, and summation is over all permutations
$\left\{ i_{1}\cdots i_{n}\right\} $ of the numbers $(1\cdots n)$
($\theta\left(x\right)$ is the Heaviside step function). Indeed the
Ward identities (\ref{ward}) are satisfied, as can be shown explicitly
by deriving Eq. (\ref{eq:wardOperators}). For instance for $n=3$,
denoting $\delta_{ij}=\delta\left(t_{i}-t_{j}\right)$ and $\theta_{ij}=\theta\left(t_{i}-t_{j}\right)$,
and using that $\partial/\partial x\ \theta\left(\pm x\right)=\pm\delta\left(\pm x\right)$,
\begin{equation*}
\begin{split}\partial_{t_{1}}\braket{i|\hat{l}_{a_{1}}\left(t_{1}\right)\hat{l}_{a_{2}}\left(t_{2}\right)\hat{l}_{a_{3}}\left(t_{3}\right)|j}= & \delta_{12}\left(\theta_{23}\braket{i|\left[\hat{l}_{a_{1}},\hat{l}_{a_{2}}\right]\hat{l}_{a_{3}}|j}+\theta_{32}\braket{i|\hat{l}_{a_{3}}\left[\hat{l}_{a_{1}},\hat{l}_{a_{2}}\right]|j}\right)\\
 & +\delta_{13}\left(\theta_{32}\braket{i|\left[\hat{l}_{a_{1}},\hat{l}_{a_{3}}\right]\hat{l}_{a_{2}}|j}+\theta_{23}\braket{i|\hat{l}_{a_{2}}\left[\hat{l}_{a_{1}},\hat{l}_{a_{3}}\right]|j}\right)\\
= & if_{a_{1}a_{2}}^{\ \ b}\delta_{12}<i|\hat{l}_{b}\left(t_{2}\right)\hat{l}_{a_{3}}\left(t_{3}\right)|j>+if_{a_{1}a_{3}}^{\ \ b}\delta_{13}<i|\hat{l}_{a_{2}}\left(t_{2}\right)\hat{l}_{b}\left(t_{3}\right)|j>,
\end{split}
\end{equation*}
where we used the ETC (\ref{EqualTimeCommutator}), and in general
(\ref{ward}) are verified. One can also check that the BJL limit
is satisfied by (\ref{eq:wardOperators}). Indeed (see~\cite{jackiw})
using the integral representation of the step function
\begin{equation}
\theta\left(t\right)=\frac{i}{2\pi}\int d\alpha e^{-i\alpha t}\frac{1}{\alpha+i\epsilon},
\end{equation}
and (\ref{eq:wardOperators}) for $n=2$, Eq. (\ref{BJL}) can be
rewritten as
\begin{equation}
\begin{split} & \frac{i}{2\pi}\lim_{p_{0}\rightarrow\infty}p_{0}\int dt\int\frac{d\alpha}{\alpha+i\epsilon}e^{ip_{0}\left(t-t_{1}\right)}\left(e^{-i\alpha\left(t-t_{1}\right)}\braket{i|\hat{l}_{a}\hat{l}_{a_{1}}|j}+e^{i\alpha\left(t-t_{1}\right)}\braket{i|\hat{l}_{a_{1}}\hat{l}_{a}|j}\right)\\
= & \frac{i}{2\pi}\lim_{p_{0}\rightarrow\infty}p_{0}\int dt\int dp_{0}'e^{ip_{0}'\left(t-t_{1}\right)}\left(\frac{1}{p_{0}-p_{0}'+i\epsilon}\braket{i|\hat{l}_{a}\hat{l}_{a_{1}}|j}-\frac{1}{p_{0}-p_{0}'-i\epsilon}\braket{i|\hat{l}_{a_{1}}\hat{l}_{a}|j}\right)
\end{split}
\end{equation}
Taking the limit for $p_{0}\rightarrow\infty$ it becomes
\begin{equation}
i\int dt\delta\left(t-t_{1}\right)\braket{i|\left[\hat{l}_{a},\hat{l}_{a_{1}}\right]|j}=i\braket{i|\left[\hat{l}_{a},\hat{l}_{a_{1}}\right]\left(t_{1}\right)|j}.
\end{equation}

Finally, we obtain the correlation function
\begin{equation}
\begin{split}\braket{i|\exp\left(ip_{\mu}\int_{0}^{L}\hat{\upsilon}^{\mu}\left(t\right)dt\right)|j}= & \sum_{n=0}^{\infty}\frac{i^{n}}{n!}p_{\mu_{1}}p_{\mu_{2}}\cdots p_{\mu_{n}}\prod_{k=1}^{n}\int_{0}^{L}dt_{k}\braket{i|\hat{\upsilon}^{\mu_{1}}\left(t_{1}\right)\hat{\upsilon}^{\mu_{2}}\left(t_{2}\right)\cdots\hat{\upsilon}^{\mu_{n}}\left(t_{n}\right)|j}\\
= & \sum_{n=0}^{\infty}\frac{i^{n}}{n!}p_{\mu_{1}}p_{\mu_{2}}\cdots p_{\mu_{n}}\prod_{k=1}^{n}\int_{0}^{L}dt_{k}\prod_{m=1}^{n-1}\theta\left(t_{m}-t_{m+1}\right)\sum_{i_{1}\cdots i_{n}}\braket{i|\hat{\upsilon}^{\mu_{i_{1}}}\cdots\hat{\upsilon}^{\mu_{i_{n}}}|j}\\
= & \sum_{n=0}^{\infty}\frac{i^{n}}{n!}L^{n}p_{\mu_{1}}p_{\mu_{2}}\cdots p_{\mu_{n}}\braket{i|\hat{\upsilon}^{\mu_{i_{1}}}\cdots\hat{\upsilon}^{\mu_{i_{n}}}|j}\\
= & \braket{i|\exp\left(iLp\cdot\hat{\upsilon}\right)|j},
\end{split}
\label{eq:correlationOperators-1}
\end{equation}
where we used Eq. (\ref{eq:wardOperators}), relation
\begin{equation}
\prod_{k=1}^{n}\int_{0}^{L}dt_{k}\prod_{m=1}^{n-1}\theta\left(t_{m}-t_{m+1}\right)=\frac{L^{n}}{n!},
\end{equation}
and the fact that the indices $\mu,\nu,\rho,\dots$ are saturated
by the symmetric terms $p_{\mu}p_{\nu}p_{\rho}\cdots$, so that the
sum over permutations of indices $i_{1},i_{2},\dots i_{n}$ gives
another $n!$ factor. On the r.h.s. of (\ref{eq:correlationOperators-1})
the $\hat{\upsilon}^{\mu}$ are matrices $\pi\left(\hat{\upsilon}^{\mu}\right)$ in
the given $\mathfrak{so}\left(4,1\right)$ representation.

\section{Matrix representation for the spin 1 sector}

\label{sec:DKP}

\subsection{The relation between $\ell=1$ representation of $SO(4,1)$ and the
DKP algebra}

For $k=1$ Eq. (\ref{charactEq}) $\beta_{\mu}=-i\hat{\upsilon}_{\mu}$ gives
\begin{equation}
\beta_{\mu}^{3}=\eta_{\mu\mu}\beta_{\mu}.\label{eq:DKP1}
\end{equation}
Plugging this into (\ref{commutationAlpha}) we obtain, for $\rho=\nu$
and $\mu\neq\nu$
\[
\beta_{\mu}\beta_{\nu}^{2}-2\beta_{\nu}\beta_{\mu}\beta_{\nu}+\beta_{\nu}^{2}\beta_{\mu}=\beta_{\mu}\eta_{\nu\nu}.
\]
Multiplying from the right by $\beta_{\nu}$ and using (\ref{eq:DKP1})
follows
\begin{equation}
\beta_{\nu}^{2}\beta_{\mu}\beta_{\nu}=2\beta_{\nu}\beta_{\mu}\beta_{\nu}^{2}.\label{eq:DKP1/2}
\end{equation}
Multiplying from the right and left by $\beta_{\nu}$ follow respectively
\begin{equation}
\left.\begin{gathered}\beta_{\nu}^{2}\beta_{\mu}\beta_{\nu}^{2}=2\beta_{\nu}\beta_{\mu}\beta_{\nu}\eta_{\nu\nu}\\
\beta_{\nu}\beta_{\mu}\beta_{\nu}\eta_{\nu\nu}=2\beta_{\nu}^{2}\beta_{\mu}\beta_{\nu}^{2}
\end{gathered}
\right\} \Rightarrow\beta_{\nu}\beta_{\mu}\beta_{\nu}=0\qquad\mu\neq\nu,\label{eq:DKP1/3}
\end{equation}
so that (\ref{eq:DKP1/2}) becomes
\begin{equation}
\beta_{\mu}\beta_{\nu}^{2}+\beta_{\nu}^{2}\beta_{\mu}=\beta_{\mu}\eta_{\nu\nu}\qquad\mu\neq\nu.\label{eq:DKP2}
\end{equation}
Now we take $\mu\neq\nu\neq\rho$ in (\ref{commutationAlpha}), multiply
from the left twice by $\beta_{\nu}$ and use (\ref{eq:DKP1/3}) and
(\ref{eq:DKP1}) to get
\[
\begin{gathered}\beta_{\nu}\beta_{\mu}\beta_{\rho}\eta_{\nu\nu}+\beta_{\nu}^{2}\beta_{\rho}\beta_{\mu}\beta_{\nu}=0\qquad\mu\neq\nu\neq\rho\end{gathered}
.
\]
Finally we use relation (\ref{eq:DKP2}) and obtain
\begin{equation}
\begin{gathered}\beta_{\nu}\beta_{\mu}\beta_{\rho}+\beta_{\rho}\beta_{\mu}\beta_{\nu}=0\qquad\mu\neq\nu\neq\rho\end{gathered}
.\label{eq:DKP3}
\end{equation}
Relations (\ref{eq:DKP1}), (\ref{eq:DKP2}) and (\ref{eq:DKP3})
define the DKP (Duffin-Kemmer-Petiau) algebra:
\begin{equation}
\beta_{\mu}\beta_{\rho}\beta_{\nu}+\beta_{\nu}\beta_{\rho}\beta_{\mu}=\beta_{\mu}\eta_{\nu\rho}+\beta_{\nu}\eta_{\mu\rho}.\label{eq:DKPalpha}
\end{equation}

\subsection{5-dimensional representation of DKP matrices for spin-0}

The DKP matrices (\ref{eq:DKPbeta}) admit a five dimensional irreducible
representation that carries the degrees of freedom of a spin-0 field
theory. We here report an explicit expression for the matrices in
this representation \cite{LunardiGauge} . These are given as
\begin{equation}
\beta_{0}=\left(\begin{array}{ccccc}
0 & 0 & 0 & 0 & 1\\
0 & 0 & 0 & 0 & 0\\
0 & 0 & 0 & 0 & 0\\
0 & 0 & 0 & 0 & 0\\
1 & 0 & 0 & 0 & 0
\end{array}\right),\quad\beta_{1}=\left(\begin{array}{ccccc}
0 & 0 & 0 & 0 & 0\\
0 & 0 & 0 & 0 & 1\\
0 & 0 & 0 & 0 & 0\\
0 & 0 & 0 & 0 & 0\\
0 & -1 & 0 & 0 & 0
\end{array}\right),\quad\beta_{2}=\left(\begin{array}{ccccc}
0 & 0 & 0 & 0 & 0\\
0 & 0 & 0 & 0 & 0\\
0 & 0 & 0 & 0 & 1\\
0 & 0 & 0 & 0 & 0\\
0 & 0 & -1 & 0 & 0
\end{array}\right),\quad\beta_{3}=\left(\begin{array}{ccccc}
0 & 0 & 0 & 0 & 0\\
0 & 0 & 0 & 0 & 0\\
0 & 0 & 0 & 0 & 0\\
0 & 0 & 0 & 0 & 1\\
0 & 0 & 0 & -1 & 0
\end{array}\right).\label{spin0beta}
\end{equation}
Notice that with this definition the spin-0 $\beta$ matrices coincide
with the defining representation for the generators corresponding
to the ``momentum sector'' of $SO(4,1)$, i.e.
$\beta_{\mu}\equiv M_{\mu4}$, where $\left(M_{AB}\right)_{KL}:=\eta_{AK}\delta_{KL}-\eta_{BK}\delta_{AL}$ (where here $\eta_{AB}=\text{diag}(1,-1,-1,-1,-1)$).
In this representation the projection of the field is
\begin{equation}
\psi=\left(\begin{array}{c}
\pi_{0}\\
\pi_{1}\\
\pi_{2}\\
\pi_{3}\\
\varphi
\end{array}\right)\qquad\longrightarrow\qquad\phi={\cal P}\psi=\left(\begin{array}{c}
0\\
0\\
0\\
0\\
\varphi
\end{array}\right).
\end{equation}

\subsection{10-dimensional representation of DKP matrices for spin-1}

The ten dimensional representations for the $\beta$ matrices can
be given as~\cite{umezawa}
\begin{equation}
\begin{gathered}\beta_{0}=\left(\begin{array}{cccccccccc}
0 & 0 & 0 & 0 & 0 & 0 & 1 & 0 & 0 & 0\\
0 & 0 & 0 & 0 & 0 & 0 & 0 & 1 & 0 & 0\\
0 & 0 & 0 & 0 & 0 & 0 & 0 & 0 & 1 & 0\\
0 & 0 & 0 & 0 & 0 & 0 & 0 & 0 & 0 & 0\\
0 & 0 & 0 & 0 & 0 & 0 & 0 & 0 & 0 & 0\\
0 & 0 & 0 & 0 & 0 & 0 & 0 & 0 & 0 & 0\\
1 & 0 & 0 & 0 & 0 & 0 & 0 & 0 & 0 & 0\\
0 & 1 & 0 & 0 & 0 & 0 & 0 & 0 & 0 & 0\\
0 & 0 & 1 & 0 & 0 & 0 & 0 & 0 & 0 & 0\\
0 & 0 & 0 & 0 & 0 & 0 & 0 & 0 & 0 & 0
\end{array}\right),\quad\beta_{1}=\left(\begin{array}{cccccccccc}
0 & 0 & 0 & 0 & 0 & 0 & 0 & 0 & 0 & 1\\
0 & 0 & 0 & 0 & 0 & 0 & 0 & 0 & 0 & 0\\
0 & 0 & 0 & 0 & 0 & 0 & 0 & 0 & 0 & 0\\
0 & 0 & 0 & 0 & 0 & 0 & 0 & 0 & 0 & 0\\
0 & 0 & 0 & 0 & 0 & 0 & 0 & 0 & 1 & 0\\
0 & 0 & 0 & 0 & 0 & 0 & 0 & -1 & 0 & 0\\
0 & 0 & 0 & 0 & 0 & 0 & 0 & 0 & 0 & 0\\
0 & 0 & 0 & 0 & 0 & 1 & 0 & 0 & 0 & 0\\
0 & 0 & 0 & 0 & -1 & 0 & 0 & 0 & 0 & 0\\
-1 & 0 & 0 & 0 & 0 & 0 & 0 & 0 & 0 & 0
\end{array}\right),\\
\beta_{2}=\left(\begin{array}{cccccccccc}
0 & 0 & 0 & 0 & 0 & 0 & 0 & 0 & 0 & 0\\
0 & 0 & 0 & 0 & 0 & 0 & 0 & 0 & 0 & 1\\
0 & 0 & 0 & 0 & 0 & 0 & 0 & 0 & 0 & 0\\
0 & 0 & 0 & 0 & 0 & 0 & 0 & 0 & 1 & 0\\
0 & 0 & 0 & 0 & 0 & 0 & 0 & 0 & 0 & 0\\
0 & 0 & 0 & 0 & 0 & 0 & 1 & 0 & 0 & 0\\
0 & 0 & 0 & 0 & 0 & -1 & 0 & 0 & 0 & 0\\
0 & 0 & 0 & 0 & 0 & 0 & 0 & 0 & 0 & 0\\
0 & 0 & 0 & 1 & 0 & 0 & 0 & 0 & 0 & 0\\
0 & -1 & 0 & 0 & 0 & 0 & 0 & 0 & 0 & 0
\end{array}\right),\quad\beta_{3}=\left(\begin{array}{cccccccccc}
0 & 0 & 0 & 0 & 0 & 0 & 0 & 0 & 0 & 0\\
0 & 0 & 0 & 0 & 0 & 0 & 0 & 0 & 0 & 0\\
0 & 0 & 0 & 0 & 0 & 0 & 0 & 0 & 0 & 1\\
0 & 0 & 0 & 0 & 0 & 0 & 0 & 1 & 0 & 0\\
0 & 0 & 0 & 0 & 0 & 0 & 1 & 0 & 0 & 0\\
0 & 0 & 0 & 0 & 0 & 0 & 0 & 0 & 0 & 0\\
0 & 0 & 0 & 0 & 1 & 0 & 0 & 0 & 0 & 0\\
0 & 0 & 0 & -1 & 0 & 0 & 0 & 0 & 0 & 0\\
0 & 0 & 0 & 0 & 0 & 0 & 0 & 0 & 0 & 0\\
0 & 0 & -1 & 0 & 0 & 0 & 0 & 0 & 0 & 0
\end{array}\right).
\end{gathered}
\end{equation}
with these representations the field $\psi$ is projected as
\begin{equation}
\psi=\left(\begin{array}{c}
F_{23}\\
F_{31}\\
F_{12}\\
F_{01}\\
F_{02}\\
F_{03}\\
A_{1}\\
A_{2}\\
A_{3}\\
A_{0}
\end{array}\right)\qquad\longrightarrow\qquad{\cal A}_{\mu}={\cal R}_{\mu}\psi=\left(\begin{array}{c}
0\\
0\\
0\\
0\\
0\\
0\\
0\\
0\\
0\\
A_{\mu}
\end{array}\right),
\end{equation}
while
\begin{equation}
\bar{{\cal A}}_{\mu}=-\psi^{\dagger}\eta_{0}{\cal R}_{\mu}^{\dagger}=\left(\begin{array}{cccccccccc}
0 & 0 & 0 & 0 & 0 & 0 & 0 & 0 & 0 & \eta_{\mu\mu}{\cal A}_{\mu}\end{array}\right)
\end{equation}
so that the scalar product is given by
\begin{equation}
\sum_{\mu}\bar{{\cal A}}_{\mu}{\cal A}_{\mu}={\cal A}_{\mu}{\cal A}^{\mu}=\eta^{\mu\nu}A_{\mu}A_{\nu}=A_{\mu}A^{\mu}
\end{equation}

\end{document}